\documentclass[11pt,twoside]{article}

\usepackage{asp2004}
\usepackage{epsf}
\usepackage{psfig}
\usepackage{amssymb}

\begin{document}
\title{The Local Interstellar Medium}   %%% Fill in title
\author{Seth Redfield}   %%% Fill in author names
\affil{Department of Astronomy and McDonald Observatory, University of Texas, Austin, TX, USA}    %%% Fill in author affiliations

\begin{abstract} %%% Abstract to run on from here.
The Local Interstellar Medium (LISM) is a unique environment that
presents an opportunity to study general interstellar phenomena in
great detail and in three dimensions.  In particular, high resolution
optical and ultraviolet spectroscopy have proven to be powerful tools
for addressing fundamental questions concerning the physical
conditions and three-dimensional (3D) morphology of this local
material.  After reviewing our current understanding of the structure
of gas in the solar neighborhood, I will discuss the influence that
the LISM can have on stellar and planetary systems, including LISM
dust deposition onto planetary atmospheres and the modulation of
galactic cosmic rays through the astrosphere --- the balancing
interface between the outward pressure of the magnetized stellar wind
and the inward pressure of the surrounding interstellar medium.  On
Earth, galactic cosmic rays may play a role as contributors to ozone
layer chemistry, planetary electrical discharge frequency, biological
mutation rates, and climate.  Since the LISM shares the same volume as
practically all known extrasolar planets, the prototypical debris
disks systems, and nearby low-mass star-formation sites, it will be
important to understand the structures of the LISM and how they may
influence planetary atmospheres.

\end{abstract}

%%% MAIN BODY OF TEXT GOES HERE. CONSULT "INSTRUCTIONS FOR AUTHORS USING
%%% LATEX2E MARKUP", SECTIONS 2.3-2.6 FOR HELP WITH EQUATIONS, FIGURES,
%%% AND TABLES.

%\section{}   %%% Top level section head (remove "%" symbol)
%\subsection{}   %%% Second level section head (remove "%" symbol)
%\subsubsection{}   %%% Lowest level section head (remove "%" symbol)
%\section*{}	%%% Unnumbered top level section head (remove "%" symbol)
%\subsection*{}   %%% Unnumbered second level section head (remove "%" symbol)

\section{Introduction}

The interstellar medium (ISM) is a critical component of galactic
structure.  Its role in the lifecycle of stars, mediating the
transition from stellar death to stellar birth, evokes a sense of a
``galactic ecology'' \citep{burton04}.  The ISM provides a platform
for the recycling of stellar material, by transferring and mixing the
remnants of stellar nucleosynthesis and creating environments
conducive for the creation of future generations of stars and planets.
It also transfers energy and momentum, absorbing flows from supernovae
blasts and strong winds from young stars and coupling these peculiar
motions with galactic rotation and turbulence.  Ultimately, when a
dense interstellar cloud collapses, it is the conservation of momentum
from the parent cloud that leads to the formation of a protostellar
disk from which stars and planets are formed.

The local interstellar medium (LISM) is the interstellar material that
resides in close ($\lesssim$100\,pc) proximity to the Sun.  For a
discussion on more distant ISM structures, see McClure-Griffiths, in
this volume.  Proximity is a special characteristic that drives much
of the interest in the LISM.  First, proximity provides an opportunity
to observe general ISM phenomena in great detail, and in three
dimensions.  ISM structures and processes are repeated almost {\it ad
infinitum} in our own galaxy \citep{dickey90}, and beyond in other
galaxies \citep{mccray87}, even at high redshift \citep{rauch99}.
Knowledge of general ISM phenomena in our local corner of the galaxy,
discussed in \S2, can be applied to more distant and difficult to
observe parts of the universe.

Second, proximity implies an interconnectedness.  The relationship
between stars and their surrounding interstellar environment will be
discussed in \S3, with particular attention paid to the interaction of
the Sun with the LISM.  In \S4, the consequences of the relationship
between stars and the LISM on planetary atmospheres are discussed, and
the LISM-Earth, or more generally, the ISM-planet connection is
explored in more detail.

This manuscript should not be considered a comprehensive review of the
subject of the LISM, but an individual, and biased, thread through a
rich research area.  I certainly will not be able to explore many LISM
topics to the level they deserve, nor will I be able to highlight all
the work done by the large number of researchers in this area.
Hopefully, this short review will introduce you to some new ideas, and
the references provided can escort you to even more work that was not
specifically mentioned in this manuscript.

\section{Properties of the Local Interstellar Medium\label{sec:part1}}

Measuring the morphological and physical characteristics of the
nearest interstellar gas has long been of interest to astronomers.
Observations of the general ISM via interstellar extinction of
background stars \citep{neckel80}, the 21\,cm H\,{\scriptsize{I}}
hyperfine transition \citep{lockman02}, or foreground interstellar
absorption in optical resonance lines \citep{cowie86} typically focus
on more distant ISM environments due to the observational challenges
inherent in measuring the properties of the LISM (see \S2.2).  Recent
reviews that focus specifically on the LISM by \citet{ferlet99} and
\citet{frisch95,frisch04} also provide some discussion on the history
of the field.  

  \subsection{Our Cosmic Neighborhood}

\begin{figure}[!ht]
%\epsscale{1.0}
%\plotone{redfieldfig1.eps}
%\plotfiddle{try4.eps}{155pt}{0}{50}{50}{-170}{-17}
\plotfiddle{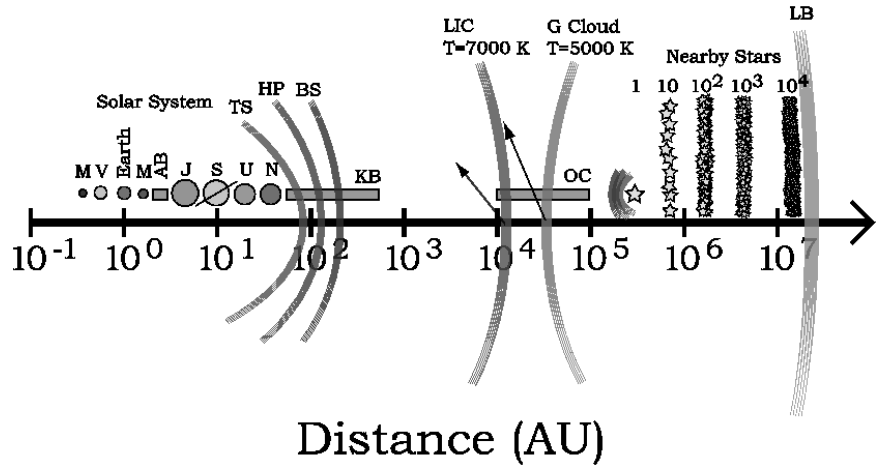}{154pt}{0}{60}{60}{-184}{-161}
\caption{Our cosmic neighborhood, shown on a logarithmic scale.  The
Solar System includes the major planets, the asteroid belt (AB), the
Kuiper belt (KB), and the Oort cloud (OC).  As of October 2005,
Voyager 1 was 97.0\,AU from the Sun.  The heliospheric structure
consists of the termination shock (TS), the heliopause (HP), and the
bow shock (BS).  The Local Interstellar Cloud (LIC) is the warm,
partially ionized cloud that directly surrounds our solar system and
currently determines the size and shape of the heliosphere.  Our
nearest neighboring cloud is the Galactic (G) Cloud which directly
surrounds the $\alpha$\,Cen stellar system.  Analogous to the
heliosphere, the $\alpha$\,Cen system includes an astrosphere.  The
LISM resides in a larger ISM structure, the Local Bubble (LB).  Within
the Local Bubble, which extends $\sim$100\,pc from the Sun, lie the
nearest $10^4$--$10^5$ stars.  This figure is inspired by a diagram in
\citet{mewaldt01}.\label{fig1}}
\end{figure}

The LISM is a diverse collection of gas.  The outer bound of what I
consider to be ``local'' in the context of the LISM, is the edge of
Local Bubble (LB), coincident, by definition, with the location of the
nearby dense molecular clouds, such as Taurus and Ophiuchus
\citep{lallement03}.  Figure\,\ref{fig1} is a schematic illustration
of the volume populated by the LISM, adapted from a
similar figure by \citet{mewaldt01}.  Within the LISM volume, our
cosmic neighborhood so to speak, resides the nearest $10^4$ to $10^5$
stars, including almost all known planetary systems (see Ford, in this
volume) and the prototypical debris disk stars (see Chen, in this
volume).  Among these stars drifts interstellar gas known as the LISM.
Several warm partially ionized clouds, such as the Local Interstellar
Cloud (LIC) and the Galactic (G) Cloud, are observed within the Local
Bubble.

The LIC is the material that directly surrounds our solar system.  The
internal pressure and ram pressure of the LIC, functions of its
density and velocity relative to the Sun, balance the force of the
outward-moving solar wind to define the boundary and shape of the
heliosphere (see \S3.1).  The heliospheric structure is not unique to
the Sun, but is observed around other stars, including the nearest
stellar system, $\alpha$\,Cen \citep{wood01}.  Nor is the heliospheric
structure static.  The heliosphere will expand and contract, in
response to the density of the LISM material surrounding the Sun (see
\S3.2).

The boundary between the LISM and the solar system is not an obvious
one.  The Oort cloud contains the most distant objects that are
gravitationally bound to the Sun, which reside at a third of the
distance to $\alpha$\,Cen.  The Oort cloud is completely enclosed by
LISM material, and due to the short extent of the LIC in the direction
of $\alpha$\,Cen \citep{redfield00}, Oort cloud objects in that
direction may be surrounded by a different collection of gas (G Cloud
material) than surrounds our planetary system (LIC material).
Currently, even much of the Kuiper belt (KB) extends beyond the bow
shock (BS) into pristine LISM material (see Sheppard, in this volume).
Some neutral LISM material can penetrate well into the solar system.
This material is utilized to make {\it in situ} measurements of
properties of the LIC (see \S2.2).  The interconnectedness of our
solar system with our surrounding interstellar medium could have
important consequences for planetary atmospheres (see \S4).

%    \subsubsection{Hot Gas: Local Bubble}
\vspace{1.5mm}

\noindent{\it Hot Gas: Local Bubble}\hspace{3mm}The Local Bubble
refers to the apparent lack of dense cold material within
approximately 100\,pc of the Sun.  Therefore, the distance to the edge
of the Local Bubble cavity is equal to the distance at which cold
dense gas is first observed.  \citet{lallement03} were able to map out
the contours of the Local Bubble by tracing the onset of the detection
of foreground Na\,{\scriptsize{I}} absorption lines in the spectra of
$\sim$1000 early type stars.  In general, the interstellar material
within the Local Bubble is too hot for neutral sodium, and therefore
none is detected until the edge of the Local Bubble is reached.  The
edge of the Local Bubble can be as close as 60\,pc, and as far as
$\sim$250\,pc, or even unbound, as toward the north and south galactic
poles \citep{lallement03}.
	
The hot gas within the Local Bubble is notoriously difficult to
observe.  Early direct detections of million degree gas came from
diffuse soft X-ray emission \citep{snowden98}, although part of this
emission is now thought to be caused by charge exchange reactions in
the heliosphere, the same process that causes comets to emit X-rays
\citep{lallement04}.  Absorption lines of highly ionized elements are
generally weak or not detected, as \citet{oegerle05} found when
looking for O\,{\scriptsize{VI}} absorption toward nearby white
dwarfs.  Detecting emission from highly ionized atoms has also been
more difficult than expected.  Using the {\it Cosmic Hot Interstellar
Plasma Spectrometer} ({\it CHIPS}), \citet{hurwitz05} do not detect
the array of extreme ultraviolet emission lines that are predicted
from the ``standard'' Local Bubble temperature and density.
Canonically, it is thought that the Local Bubble is filled with
$T\,\sim\,10^6$\,K, $n\,\sim\,5\,\times\,10^{-3}$\,cm$^{-3}$ gas that
extends about $R\,\sim\,100\,$pc.  However, much work remains to be
done to understand the nature of the hot Local Bubble gas.

%Theoretical explanations for the origin and evolution of the Local
%Bubble are still controversial, and beyond the scope of
%this short review.  More information can be found in reviews by
%\citet{breitschwerdt04} and \citet{cox05}.

%    \subsubsection{Warm Gas: Local Interstellar Clouds}
\vspace{1.5mm}

\noindent{\it Warm Gas: Local Interstellar Clouds}\hspace{3mm}Within the hot Local Bubble substrate are dozens of individual
accumulations of diffuse gas that are warm and partially ionized
($T\,\sim\,7000$\,K, $n\,\sim\,0.3$\,cm$^{-3}$,
$R\,\sim\,0.5$--$5\,$pc).  It is most commonly this material that is
being referenced with the term ``local interstellar medium.''  
%It has
%also been referred to as the ``local fluff'' (LF), ``complex of local
%interstellar clouds'' (CLIC), ``very local interstellar medium''
%(VLISM).  
It is warm, partially ionized material that directly surrounds our
solar system, and which can be measured with {\it in situ}
observations and high resolution optical and ultraviolet (UV)
spectroscopy (see \S2.2).  The warm LISM will dominate the remainder
of this review, because it is the best studied of the different phases
of LISM material, and the most significant with regards to interaction
with stars.

\begin{table}[!ht] 
\caption{Properties of Warm Local Interstellar Clouds} 
\smallskip 
\begin{center} 
{\scriptsize \begin{tabular}{lccl} 
\tableline 
\noalign{\smallskip} 
Property & Value & Ref. & Comments$^{\rm a}$ \\
\noalign{\smallskip} 
\tableline 
\noalign{\smallskip} 
Temperature ($T$)  & 6680 $\pm$ 1490 K & 1 & LISM, AL \\
               & 6300 $\pm$ 340 K & 2 & LIC, {\it IS} \\
Turbulent Velocity ($\xi$) & 2.24 $\pm$ 1.03 km s$^{-1}$ & 1 & LISM, AL \\
Velocity Magnitude ($v_0$) & 25.7 $\pm$ 0.5 km s$^{-1}$ & 3 & LIC, AL \\
               & 28.1 $\pm$ 4.6 km s$^{-1}$ & 4 & LISM, AL \\
               & 26.3 $\pm$ 0.4 km s$^{-1}$ & 2 & LIC, {\it IS} \\
Velocity Direction ($l_0$, $b_0$) & 186$^{\circ}\hspace{-.95ex}.$1, --16$^{\circ}\hspace{-.95ex}.$4 & 3 & LIC, AL \\
               & 192$^{\circ}\hspace{-.95ex}.$4, --11$^{\circ}\hspace{-.95ex}.$6 & 4 & LISM, AL \\
               & 183$^{\circ}\hspace{-.95ex}.$3 $\pm$ 0$^{\circ}\hspace{-.95ex}.$5, --15$^{\circ}\hspace{-.95ex}.$9 $\pm$ 0$^{\circ}\hspace{-.95ex}.$2 & 2 & LIC, {\it IS} \\
H{\tiny{I}} Column Density ($\log N_{\rm HI}$) & 17.18 $\pm$ 0.70 cm$^{-2}$ & 5 & LISM, AL \\
He{\tiny{I}} Volume Density ($n_{\rm HeI}$) & 0.0151 $\pm$ 0.0015 cm$^{-3}$ & 6 & LIC, {\it IS} \\
H{\tiny{I}} and He{\tiny{I}} Ratio ($N_{\rm HI}/N_{\rm HeI}$) & 14.7 $\pm$ 2.0 & 7 & LISM, AL \\
Electron Volume Density ($n_e$) & 0.11$^{+0.12}_{-0.06}$ cm$^{-3}$ & 8 & LIC, AL \\
H{\tiny{I}} Volume Density ($n_{\rm HI}$) & 0.222 $\pm$ 0.037 cm$^{-3}$ & 6,7 & $n_{\rm HI}\,=\,n_{\rm HeI}\,\times\,N_{\rm HI}/N_{\rm HeI}$ \\
Thermal Pressure ($P_T/k$) & 3180$^{+1850}_{-1130}$ K cm$^{-3}$ & 1,6,7 & $P_T\,=\,nkT$ \\
Turbulent Pressure ($P_{\xi}/k$) & 140$^{+140}_{-130}$ K cm$^{-3}$ & 1,6,7 & $P_{\xi}\,=\,0.5\rho{\xi}^2$ \\
Hydrogen Ionization ($X_{\rm H}$) & 0.33$^{+0.24}_{-0.13}$ & 6,7,8 & $X_{\rm H}\,=\,n_{\rm HII}/(n_{\rm HI}\,+\,n_{\rm HII})$ \\
Absorbers per sightline & 1.8 & 9 & LISM, AL\\
Cloud size ($r$) & $\sim$2.3 pc & 10 & LIC, AL\\
Cloud mass ($M$) & $\sim$0.32 $M_{\odot}$ & 10 & LIC, AL \\
\noalign{\smallskip} 
\tableline 
\end{tabular} 
}
\end{center} 
\vspace{-4mm}
{\scriptsize \hspace{5mm}$^{\rm a}$\,LISM\,=\,average of several LISM sightlines; LIC\,=\,quantity for LIC material only; AL\,=\,derived from absorption line observations; {\it IS}\,=\,derived from {\it in situ} observations.}\\
{\scriptsize {\hspace*{5mm}{\scshape{References.}}---(1) \citeauthor{redfield04tt} \citeyear{redfield04tt}; (2) \citeauthor{witte04} \citeyear{witte04}; (3) \citeauthor{lallement92} \citeyear{lallement92}; (4) \citeauthor*{frisch02} \citeyear{frisch02}; (5) \citeauthor{redfield04sw} \citeyear{redfield04sw}; (6) \citeauthor{gloeckler04} \citeyear{gloeckler04}; (7) \citeauthor{dupuis95} \citeyear{dupuis95}; (8) \citeauthor{wood97} \citeyear{wood97}; (9) \citeauthor{redfield02} \citeyear{redfield02}; (10) \citeauthor{redfield00} \citeyear{redfield00}; }}
\vspace{-3mm}
\end{table}

Typical properties of the local warm interstellar clouds are given in
Table\,1.  Absorption line spectroscopy and {\it in situ} observations
often provide independent measurements of the same quantity (e.g.,
$T$, $v_0$, $l_0$, $b_0$).  In addition, the two techniques are often
complementary, as when parameters derived from absorption spectra of
nearby stars (e.g., $N_{\rm HI}/N_{\rm HeI}$) are combined with {\it
in situ} measurements (e.g., $n_{\rm HeI}$) to determine a third
physical quantity that would be difficult or impossible to determine
using either technique alone (e.g., $n_{\rm HI}$).  The LISM is a
diverse collection of gas.  For example, individual temperature
determinations of LISM material can be made to the precision of
$\pm$200\,K, although temperatures ranging from 2000--11000\,K are
observed \citep{redfield04tt}.  The value given for the LISM in
Table\,1 is a weighted LISM mean.

Despite the diversity, there are several observational clues that
indicate a coherence in the LISM.  First, LISM absorption features in
high resolution spectra can almost always be fit by one to three
individual, symmetric, well-separated, Gaussian profiles, as opposed
to a broad asymmetric feature that would result from multiple
absorption features with a gradient of velocities.  Second, the
projected velocities of LISM features toward nearby stars can be
characterized by a single bulk flow \citep{lallement92,frisch02} that
matches the observed ISM flow into our solar system \citep{witte04}.
However, small deviations from this bulk velocity vector are observed
in the direction of the leading edge of the LIC, where the gas appears
to be decelerated, possibly due to a collision of LIC material with
neighboring LISM material \citep{redfield01}.  Third, chaotic small
scale structure in the LISM has not been detected.  In one example, a
collection of 18 Hyades stars, separated by only
1$^{\circ}$--10$^{\circ}$, shows a smooth slowly varying gradient in
column density with angular distance, as opposed to a chaotic,
filamentary geometry.  However, an extensive database of observations
is required to fully study the three dimensional structure of the LISM
(see \S2.3).

Several properties of warm LISM clouds are not well known.  Except for
the material currently streaming into the solar system, volume
densities are very difficult to measure.  Inherent in absorption line
observations are: (1) the ignorance of a length scale to the absorbing
material, other than the limit set by the distance of the background
star, and (2) the ignorance of density variations within a single
collection of gas, since only the total column density is observed.

Measurements of the magnetic field in the LISM have also been
difficult to make.  For lack of better measurements, observations of
distant (several kpc) pulsars give a ``local'' galactic magnetic field
strength of $\sim$1.4\,$\mu$G \citep{rand94}, although this value may
have little to do with the magnetic fields entrained in the LISM.
Polarization measurements of nearby ($<$35 pc) stars are weak, but
seem to indicate an orientation parallel to the galactic plane
\citep{tinbergen82,frisch04}.  The same orientation is derived from
heliospheric observations of 1.78--3.11\,kHz radio emission by
Voyager\,1 \citep{kurth03}.  The orientation and strength of the local
magnetic field will have important consequences on the structure and
shape of the heliosphere
\citep[e.g.,][]{gloeckler97,pogorelov04,florinski04,lallement05}.
Additionally, since the magnetic pressure goes as $B^2$, the strength
of the LISM magnetic field ($B$) could have important consequences for
the relationship between the hot Local Bubble gas and the warm LISM
clouds.  The apparent pressure imbalance between the warm
($P_{tot}/k\,\sim\,3300$\,K cm$^{-3}$, see Table\,1) and hot
($P_T/k\,\sim\,10000$\,K cm$^{-3}$, see ``Hot Gas'' section above)
components of the LISM has been a persistent topic concerning the
structure of our local interstellar environment \citep{jenkins02}.
Possible scenarios include: (1) the presence of a LISM magnetic field
of $\sim$4.8\,$\mu$G to match the hot Local Bubble pressure, (2)
refinement of Local Bubble observations and models that may reduce the
Local Bubble pressure, in particular refining the X-ray and extreme
ultraviolet observations of nearby hot gas
\citep{hurwitz05,lallement04}, or (3) the hot and warm components of
the LISM are not in pressure equilibrium.  It will be important to
resolve this issue in order to understand the interaction between the
hot Local Bubble gas and warm local interstellar clouds.

%    \subsubsection{Cold Gas?}
\vspace{1.5mm}

\noindent{\it Cold Gas?}\hspace{3mm}Although the volume of the Local Bubble is defined by the scarcity of
cold dense gas, there are some indications that collections of
cold gas may reside within the Local Bubble.  Observations of
Na\,{\scriptsize{I}} by \citet{lallement03}, which define the
morphology of the Local Bubble, also identify a number of isolated
dense clouds, just inside the Local Bubble boundary.
\citet{magnani96} identify several small molecular clouds that are
thought to be relatively nearby ($\leq$200\,pc), although their
precise distances are not often known.  These clouds have
$T\,\sim\,20$\,K, $n\,\sim\,30$\,cm$^{-3}$, and sizes of about
$R\,\sim\,1.4\,$pc. One such nearby molecular cloud, MBM\,40
($\sim$100\,pc), contains molecular cores, although they are not
massive enough for star formation \citep{cholminh03}.  MBM\,40 is
likely an example of a dense cloud that resides within the Local
Bubble boundary.

  \subsection{How do we Measure the Properties of the LISM? \label{meas}}

The proximity of the LISM presents unique challenges and opportunities
for measuring the properties of nearby interstellar gas, which is too
sparse to cause measurable reddening or be detected in atomic hydrogen
21\,cm emission.  One unique observational technique is {\it in situ}
measurements of ISM particles, which stream directly into the inner
solar system.  These observations complement the traditional
ISM observational technique of high resolution absorption line
spectroscopy.  Observing the closest ISM provides many advantages,
such as simple absorption spectra, well known distances to
background stars, and large projected areas that allow multiple
observations through different parts of a single cloud, enabling a
probe of its properties in three dimensions.  
%Below, we explore some
%of the different observational techniques used to study the LISM.

%    \subsubsection{{\it In Situ Measurements}\label{sec:part1a}}
\vspace{1.5mm}

\noindent{\it In Situ Measurements}\hspace{3mm}A powerful
observational technique, absent in the vast majority of astrophysical
research, is the ability to send instruments to\,physically interact
with, collect, and measure the properties of the material of interest
directly, instead of relying on photons.  Due to the close proximity
of the LISM, interstellar particles are continually streaming into the
interplanetary medium.  Neutral helium atoms, and helium ``pick-up''
ions (neutral helium atoms ionized as they approach the Sun and are
``picked-up,'' or entrained, in the solar wind plasma), are observed
by mass spectrometers onboard the {\it Ulysses} spacecraft.
\citet{mobius04} review these measurements, together with
He\,{\scriptsize{I}} UV-backscattered emission, collected and analyzed
by several groups over many years.  These observations give consistent
measurements of the temperature, velocity, and density of the
interstellar medium directly surrounding the solar system (see
Table\,1).  LISM dust particles collected in the interplanetary medium
are among the sample onboard the {\it Stardust} mission, scheduled to
return to Earth in January 2006 \citep{brownlee03}.  Laboratories will
be able to analyze the collected particles in detail.  Not only will
this provide information about the nature of dust in the LISM, but may
answer questions about the origin of ISM dust and its role in
circumstellar and disk environments (see Chen, in this volume).  Both
Voyager spacecraft, launched in 1977, are still functioning and
returning data as part of the Voyager Interstellar Mission (VIM).  On
16 December 2004, Voyager\,1 provided a long sought-after measurement
of the distance to the termination shock, at 94.01\,AU.  With enough
power to last until 2020, Voyager\,1 should provide measurements of
its encounter with pristine interstellar material once it crosses the
heliopause around 2015 \citep{stone05}.

%    \subsubsection{High Resolution Absorption Line Spectroscopy\label{sec:part1b}}
\vspace{1.5mm}

%\begin{figure}[!ht]
%%\epsscale{1.0}
%%\plotone{redfieldfig2.eps}
%\plotfiddle{redfieldfig2.eps}{160pt}{0}{60}{60}{-152}{-30}
%\caption{The 100 strongest resonance lines, ranked in order of their
%hydrogen column density sensitivity.  White bars indicate those that
%fall in the optical (3000-10000\AA), gray for ultraviolet
%(1200-3000\AA), and black for far-ultraviolet (900-1200\AA).  The top
%15 lines are H\,{\scriptsize{I}} transitions from Lyman-$\alpha$ to
%Lyman-$o$, only Lyman series lines to Lyman-$\omega$ are shown.  The
%vertical dashed lines indicate the typical range of hydrogen column
%densities observed for warm LISM clouds.  Those transitions with
%sensitivities left of this range will be optically-thick and
%saturated, whereas those transitions to the right will not be
%sensitive enough to detect absorption from warm LISM
%clouds. \label{fig2}}
%\end{figure}

\begin{figure}[!t]
%\epsscale{1.0}
%\plotone{redfieldfig2.eps}
\plotfiddle{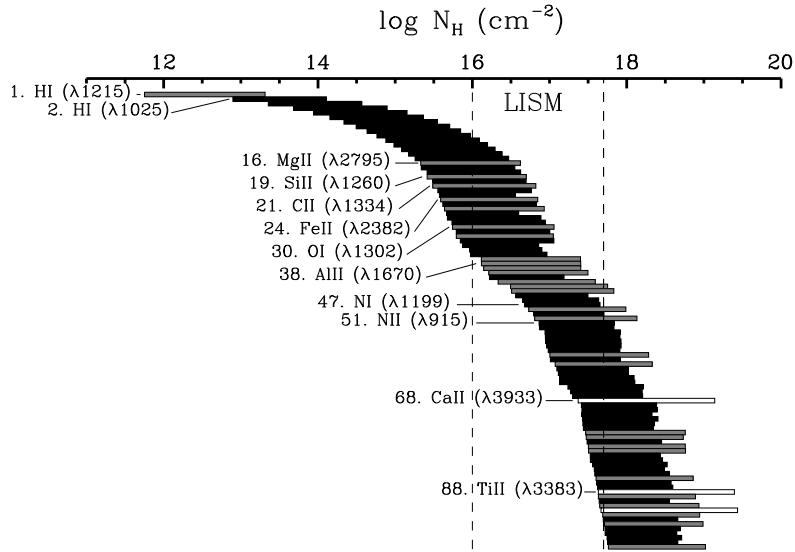}{188pt}{0}{62}{62}{-158}{-32}
\caption{The 100 strongest resonance lines, ranked in order of their
hydrogen column density sensitivity.  White bars indicate those that
fall in the optical (3000-10000\AA), gray for ultraviolet
(1200-3000\AA), and black for far-ultraviolet (900-1200\AA).  The top
15 lines are H\,{\scriptsize{I}} transitions from Lyman-$\alpha$ to
Lyman-$o$, only Lyman series lines to Lyman-$\omega$ are shown.  The
vertical dashed lines indicate the typical range of hydrogen column
densities observed for warm LISM clouds.  Those transitions with
sensitivities left of this range will be optically-thick and
saturated, whereas those transitions to the right will not be
sensitive enough to detect absorption from warm LISM
clouds. \label{fig2}}
\end{figure}

\noindent{\it High Resolution Absorption Line
Spectroscopy}\hspace{3mm}A standard technique used to measure the
physical properties of foreground interstellar material along the line
of sight toward a background star is high resolution absorption line
spectroscopy.  This kind of work has a long and rich history, but has
typically been dominated by more distant ISM environments, with large
column densities and strong absorption signatures
\citep{cowie86,savage96}.  The challenge inherent in absorption line
spectroscopy of the LISM is the low column density along sightlines to
nearby stars.  This limits the number of diagnostic lines to only the
strongest resonance line transitions.  In Figure\,\ref{fig2}, the
hydrogen column density sensitivities are shown for the 100 strongest
ground-state transitions at wavelengths from the far-ultraviolet
(FUV), through the ultraviolet (UV), to the optical.  The lower
sensitivity limit indicates a 3$\sigma$ detection in a high
signal-to-noise observation with modern high resolution instruments.
The upper sensitivity limit marks the column density at which the
transition becomes optically thick and leaves the linear part of the
curve of growth, where, although absorption is detected, limited
information can be obtained from the saturated absorption profile.
Ionization structure typical for warm LISM clouds is incorporated
\citep{slavin02,wood02}, although typical LISM depletion is not; only
solar abundances are assumed \citep{asplund05}.  The range of LISM
absorbers ($16.0\,\leq\,\log\,N_{\rm H}\,$(cm$^{-2}$)$\,\leq\,17.7$)
is such that less than 100 transitions are available to study the
LISM.  Taking into account such issues as blending or continuum
placement, which can limit the diagnostic value of an individual
transition, reduces the number of useful transitions even more.  Most
of the transitions lie in the FUV and UV, with only a few transitions
available in the optical, most importantly Ca\,{\scriptsize{II}}
resonance lines at $\sim$3950\,\AA.  (Other notable optical
transitions, such as Na\,{\scriptsize{I}} and K\,{\scriptsize{I}},
probe more distant and higher column density ISM environments, see
\citeauthor{lallement03} \citeyear{lallement03} and
\citeauthor{welty01} \citeyear{welty01}.)  Recent LISM absorption line
observations of these transitions have been made in the FUV
%(D\,{\scriptsize{I}}, C\,{\scriptsize{II}},
%C\,{\scriptsize{III}}, N\,{\scriptsize{I}}, N\,{\scriptsize{II}},
%O\,{\scriptsize{I}}, Si\,{\scriptsize{II}}, P\,{\scriptsize{II}},
%Ar\,{\scriptsize{I}}, Fe\,{\scriptsize{II}}, etc...) 
with the {\it Far Ultraviolet Spectroscopic Explorer} ({\it FUSE})
\citep[e.g.,][]{lehner03,wood02}, in the UV 
%(D\,{\scriptsize{I}},
%C\,{\scriptsize{II}}, N\,{\scriptsize{I}}, O\,{\scriptsize{I}},
%Mg\,{\scriptsize{I}}, Mg\,{\scriptsize{II}}, Al\,{\scriptsize{II}},
%Si\,{\scriptsize{II}}, Si\,{\scriptsize{III}}, Fe\,{\scriptsize{II}},
%etc...)  
with the {\it Hubble Space Telescope} ({\it HST})
\citep[e.g.,][]{redfield04sw,redfield02}, and in the optical
%(Ca\,{\scriptsize{II}}, etc...) 
with ultra-high resolution
spectrographs, such as those at McDonald Observatory and the
Anglo-Australian Observatory
\citep[e.g.,][]{crawford01,welty94}.

  \subsection{Future Directions \label{fut1}}

Absorption line analyses, supported by {\it in situ} observations of
the LIC, have resulted in numerous single sightline measurements of
projected velocity, column density, and line width for several dozens
of sightlines, leading to individual measurements of various physical
properties of the LISM, including temperature and turbulence
\citep{redfield04tt}, electron density \citep{wood97}, ionization
\citep{jenkins00}, depletion \citep{lehner03}, and small scale
structure \citep{redfield01}.  However, with the recent accumulation
of significant numbers of observations, it is possible to go beyond
the single sight\-line analysis and develop a global morphological and
physical model of the LISM.  Initial steps have been made toward this
end with global bulk flow kinematic models of the LIC and G Clouds
produced by \citet{lallement92}, and a global morphological model of
the LIC by \citet{redfield00}.

Future LISM research will synthesize the growing database of LISM
observations, taking advantage of the information contained in the
comparison of numerous individual sightlines.  A global morphological
model would enable the development of global models of various
physical properties, such as kinematics, ionization, depletion,
density, etc.  Ultimately, these global models are required to tackle
larger issues that cannot be fully addressed by single sightline
analyses, such as the interactions of clouds, the interaction of the
warm LISM clouds with the surrounding hot Local Bubble substrate, the
strength and orientation of magnetic fields, and the origin,
evolution, and ages of clouds in the LISM.  Such work will not only be
important in understanding the structure of gas in our local
environment, but will be applicable to other, more distant and
difficult to observe interstellar environments in our galaxy and
beyond.

%    \subsubsection{CLOSE: Ca\,{\scriptsize{II}} LISM Optical Survey of
%    our Environment}
\vspace{1.5mm}

\noindent{\it CLOSE: Ca\,{\scriptsize{II}} LISM Optical Survey of our
Environment}\hspace{3mm}A global morphological model of the LISM
requires high spatial and distance sampling.  The development of a
morphological model for the LIC by \citet{redfield00} was possible
because LIC absorption is detected in practically every direction,
since the material surrounds the solar system.  More distant LISM
clouds will subtend smaller angles on the sky, and will require higher
density sampling of LISM observations in order to be morphologically
characterized.  The CLOSE (Ca\,{\scriptsize{II}} LISM Optical Survey
of our Environment) project is a large scale, ultra-high resolution
survey of $\sim$500 nearby stars that will enable a global
morphological model of the LISM (work in collaboration with M.S. Sahu,
B.K. Gibson, C. Thom, A. Hughes, N. McClure-Griffiths, P. Palunas).
Previous Ca\,{\scriptsize{II}} surveys \citep[and see summary,
\citeauthor{redfield02}
\citeyear{redfield02}]{vallerga93,lallement86,lallement92,welty96}
only accumulated $\sim$50 nearby sightlines, but detected LISM
Ca\,{\scriptsize{II}} absorption in 80\% of the targets.  Our survey
will provide extensive coverage.  The maximum angular distance between
any two adjacent targets will be $<$10$^{\circ}$, including more than
45 target pairs that will be $<$1$^{\circ}$ apart, providing an
interesting study of small scale structure in the LISM.  When combined
with past and future observations, this survey will provide a
significant baseline with which to search for long-term LISM
absorption variation, as the sightlines to these high proper motion
nearby stars vary over timescales of decades.  Ultimately, the CLOSE
project will provide a valuable database for the development of a
global morphological model of the LISM.

\section{Relationship Between Stars and their Local Interstellar Medium\label{part2}}

As discussed in reference to Figure\,\ref{fig1}, the interaction
between the LISM and solar/stellar winds is mediated by the
heliospheric/astrospheric interface.  This interface is defined by the
balance between the solar/stellar wind and the LISM.  Reviews of
heliospheric modeling include \citet{zank99} and \citet{baranov90},
and the detection of astrospheres around nearby stars is reviewed by
\citet{wood04}.

  \subsection{The Heliosphere and Astrospheres\label{part2a}}

In the standard picture of the heliosphere, discussed by
\citet{zank99}, \citet{baranov90} and \citet{wood04}, the magnetized
solar wind is shocked to subsonic speeds (``termination shock''), as
is the ionized LISM material (``bow shock'').  The interface in
between (``heliopause'') is where the plasma flows of the solar wind
and LISM are deflected from each other.  It was originally thought
that neutral atoms from the LISM pass through the heliosphere
unimpeded and therefore have a negligible influence on the structure
of the heliosphere.  However, charge exchange reactions between
ionized hydrogen in the solar wind and neutral hydrogen in the LISM
act to heat and decelerate LISM hydrogen atoms just prior to the
heliopause.  The resulting structure, referred to as the ``hydrogen
wall,'' is an accumulation of hot hydrogen between the heliopause and
the bow shock.

At the same time that heliospheric simulations were indicating an
enhancement of hydrogen just beyond the heliopause, it was becoming
clear that observations of LISM absorption in H\,{\scriptsize{I}}
Lyman\,$\alpha$ were discrepant with other LISM absorption lines along
the same line of sight.  In particular, excess H\,{\scriptsize{I}}
absorption was required on the red and blue sides of the LISM
H\,{\scriptsize{I}} absorption feature, in order to be consistent with
the optically thin D\,{\scriptsize{I}} absorption profile, which is
only 82\,km\,s$^{-1}$ to the blue of the H\,{\scriptsize{I}}
absorption.  The models predicted a column density for the ``hydrogen
wall'' of $\log\,N_{\rm H}\,$(cm$^{-2}$)\,$\sim\,$14.5.  From
Figure\,\ref{fig2}, it is clear that only the Lyman series hydrogen
lines are sensitive enough to detect these low column densities.
Indeed, heliospheric H\,{\scriptsize{I}} absorption was first detected
using the Lyman-$\alpha$ profile, when \citet{linsky96} measured
excess absorption redshifted with respect to the LISM absorption.
The blueshifted excess absorption is associated with an astrosphere.
Because we observe the decelerated heliospheric hydrogen from the
inside, the heliospheric absorption is redshifted, whereas the
decelerated hydrogen in an astrosphere is observed exterior to the
astrosphere, and therefore is blueshifted, see Figure\,6 of
\citet{wood04}.

It should be noted that if the interstellar absorption gets to be too
large, $\log\,N_{\rm H}\,$(cm$^{-2}$)\,$\geq\,$18.7, the saturated ISM
absorption will obliterate any sign of a slightly offset heliospheric
or astrospheric absorption.  So, these measurements are only possible
within the low column density volume of the LISM.  Among a sample of
nearby stars, heliospheric and astrospheric absorption has been
detected for many stars \citep{wood05sup}.  Multiple heliospheric
detections sample the structure of our heliosphere in three
dimensions, while multiple astrospheric detections provide
measurements of weak solar-like winds around other stars.  More than
50\% of stars within 10\,pc that have high resolution UV
Lyman-$\alpha$ spectra show signs of astrospheric absorption
\citep{wood05let}.
%Observations of this kind are required to disentangle the
%stellar and interstellar influences on the structure of astrospheres.

  \subsection{Heliospheric Variability\label{part2b}}

%The structure of the heliosphere will vary on both short and long
%timescales.  The short term variations are dominated by solar wind
%fluctuations, while long term variations may be dominated by
%fluctuations in the surrounding ISM density.

%    \subsubsection{Short Term}
\vspace{1.5mm}

\noindent{\it Short Term}\hspace{3mm}The solar wind strength and
distribution fluctuate with the 11-year solar cycle
\citep{richardson97}.  These variations slowly propagate out to the
heliospheric boundary and it is expected that the heliopause will
expand and contract on a comparable timescale.  The stochastic injection of
energy into the solar wind in the form of flares and mass ejections
leads to variability on even shorter timescales.  This dynamic wind is
constantly buffeting the magnetospheres of planets in its path as well
as the heliospheric boundary.  Voyager\,1 may have detected such
short-term variability when over a 7-month period in 2003 the
termination shock contracted inward, over Voyager\,1, and then
expanded back outward over Voyager\,1 yet again, a year before
Voyager\,1 unambiguously crossed the termination shock
\citep{krimigis03,mcdonald03}.  Due to the immensity of ISM clouds,
even the smallest of LISM structures are not expected to contribute to
short-term ($\sim$yrs) heliospheric variability.

%    \subsubsection{Long Term}
\vspace{1.5mm}

\noindent{\it Long Term}\hspace{3mm}Long-term variations in the solar
wind strength are not well known, but observations of astrospheres
around young solar analogs provide clues as to what kind of wind
the Sun had in its distant past.  The solar wind, 3.5\,billion years ago,
may have been $\sim$35$\times$ stronger than it is
today \citep{wood05let}.  In contrast, density variations spanning 6
orders of magnitude are commonly observed throughout the general ISM.
Since the Sun has likely encountered a number of different ISM
environments with extreme variations in density, it seems quite
intuitive to expect that the variation in density of our surrounding
interstellar environment plays the dominant role in long-term
variations of the heliosphere.  Detailed models support this
intuition.  \citet{zankfrisch99} model the modern heliosphere
surrounded by a LISM density 50$\times$ the current value and find
that the termination shock shrinks from $\sim$100\,AU to $\sim$10\,AU.
In the next section, we will explore the possible consequences for
planets, such as Earth, caught in the midst of such a dramatic change
in the structure of the heliosphere.

\vspace{-.25mm}
\section{The LISM-Earth or ISM-Planet Connection\label{part3}}

Discussion of a LISM-Earth connection has an incredibly long history.
\citet{shapley21} and \citet{hoyle39} are among the earliest
references.  Research has intensified as of late; observations and
models have improved, geologic and climatic events remain unexplained,
and it is becoming clear that the habitability of planets may be
dependent on subtle astrobiological parameters
\citep[e.g.,][]{fahr68,begelman76,thaddeus86,frisch98,florinski03,yeghikyan04a,wallmann04,pavlov05,gies05}.

  \subsection{Planetary Consequences of Heliospheric Variability}

It appears inevitable that the heliosphere has and will continue to
expand and contract as the Sun passes through different ISM
environments.  The Sun is now surrounded by a relatively modest
density cloud, the LIC, and the heliopause stands well beyond the
planets at $\sim$100\,AU.  However, when the Sun encounters more dense
ISM clouds, the heliosphere will shrink, perhaps to within the inner
solar system.  The question then arises: What are the consequences, on
Earth and on other planets, of a compressed heliosphere/astrosphere?

%    \subsubsection{Cosmic Rays}
\vspace{1.5mm}

\noindent{\it Cosmic Rays}\hspace{3mm} The solar wind is magnetized
and extends out to the heliopause.  Energetic charged particles, or
cosmic rays, interact with this magnetic field and if they are not
too energetic ($\leq$\,1\,GeV), the cosmic rays are partially or
completely prevented from penetrating far into the heliosphere.  The
heliosphere is one of three screens, together with the Earth's
magnetic field and the Earth's atmosphere, that modulate the cosmic
ray flux at the surface of the Earth.  The loss of heliospheric
modulation would lead to flux increases of 10--100$\times$ at energies
of 10--100\,MeV at the top of the terrestrial magnetosphere
\citep{reedy83}, and could have serious consequences for several
planetary processes.

Cosmic rays are important sources of ionization in the upper
atmosphere, creating showers of secondary particles, ultimately
leading to muons, which dominate the cosmic ray flux in the lower
atmosphere.  Ions in the atmosphere may serve as cloud nucleation
sites, increasing low altitude clouds and ultimately increasing the
planetary albedo \citep{carslaw02}.  A connection between cosmic rays
and clouds was suggested by \citet{svensmark97} and a correlation
between cosmic ray flux and low cloud cover over the course of a solar
cycle was found by \citet{marsh00}, even though the global cosmic ray
flux varied by only $\sim$\,15\%.  The 1--2 orders of magnitude
variation that would result from the loss of the heliospheric screen
could have a tremendous impact on the formation of clouds in the
Earth's atmosphere.  The ionization caused by cosmic ray secondaries
may also trigger lightning production \citep{gurevich01}.  Cosmic rays
increase the production of NO and NO$_2$ in the upper stratosphere,
which significantly influences ozone layer chemistry.
\citet{randall05} tracked the enhancement of nitric oxide and nitrogen
dioxide that resulted from an injection of low energy cosmic rays
following a series of intense solar storms.  In some locations within
the polar vortex, the increased levels of NO and NO$_2$ led to 60\%
reductions in terrestrial ozone over several months.

Muons, electrons, and other cosmic ray products are significant
natural radiation sources, accounting for 30--40\% of the annual dose
from natural radiation in the United States \citep{alpen98}.  Muons
ionize atoms in our bodies, producing hydroxyl radicals, which can
cause DNA mutations.  The cosmic muon flux can be significant even at
depths of 1\,km below the Earth's surface, and is therefore a source
of mutation even for deep-sea or deep-earth organisms.  The loss of
the heliospheric screen, and the subsequent increase in cosmic ray
flux on the Earth's atmosphere, could have important implications for
the long-term evolution of the Earth's climate, and for long-term
mutation rates in terrestrial organisms.

%    \subsubsection{Dust Deposition}
\vspace{1.5mm}

\noindent{\it Dust Deposition}\hspace{3mm}Passage through dense ISM
environments will compress the heliosphere while also depositing a
significant amount of interstellar dust onto the top of the Earth's
atmosphere.  \citet{pavlov05} presented atmospheric models in which
large amounts of interstellar dust cause a reverse greenhouse effect,
blocking or scattering incident visible light while being transparent
to infrared thermal radiation.  \citet{mckay78} also find that
increased deposition of interstellar H$_2$ onto Earth's atmosphere
would decrease mesospheric ozone levels, decrease the mesospheric
temperature, and cause high altitude noctilucent ice clouds, which
would ultimately increase the planetary albedo.  Dust deposition could
be a natural trigger for ``snowball'' Earth episodes, periods of
$\sim$200,000 years in which the Earth is entirely glaciated
\citep{hoffman98}.

  \subsection{Future Directions}

Most of the work on the LISM-Earth or ISM-planet connection has been
theoretical.  The few observational tests of this relationship have
focused on correlating the largest ISM fluctuations with the largest
climatic or geological fluctuations.  For example, \citet{shaviv03}
claimed a correlation between the passage through spiral arms, the
glaciation period, and the cosmic ray flux derived from iron
meteorites.  The most significant hurdles that these kinds of
empirical tests must overcome are the extreme systematic errors that
plague long-term astronomical, climatic, and meteoritic timescales
that are so vital to demonstrating a convincing correlation.  Invoking
passage through spiral arms and major glaciations requires accurate
temporal calibration back more than a billion years.

Although it is intuitive to attach the largest fluctuations in the ISM
to the largest fluctuations in the climatic record, with respect to
the modulation of the cosmic ray flux, modest density clouds may have
a significant influence.  As demonstrated in the model by
\citet{zankfrisch99}, a modest density LISM has a dramatic effect on
the structure of the heliosphere.  Although higher density
environments would compress the heliosphere still further, there will
be a point of diminishing returns as the incident cosmic ray flux on
Earth's atmosphere approaches the galactic cosmic ray flux level.
Future work will explore the evolution of the heliosphere through a
variety of ISM environments, besides our current LISM surroundings and
intermediate to the extremes of giant molecular clouds in the heart of
spiral arms.

%    \subsubsection{Historical Solar Trajectory}
\vspace{1.5mm}

\noindent{\it Historical Solar Trajectory}\hspace{3mm}Linking the
long-term timing of ice ages with spiral arm passages will continue to
be an interesting test of the ISM-planet connection, but the LISM
provides a provocative alternative empirical test (work in
collaboration with J. Scalo and D.S. Smith).  As discussed in \S2, we
know precisely the nature of the LISM that directly surrounds our
solar system now from {\it in situ} measurements of the LIC.  If we
look at a very nearby star, in the direction of the historical solar
trajectory \citep{dehnen98}, the observed LISM absorption should
provide information on the nature of the LISM that the Sun encountered
only a short time ago. (The Sun travels 1\,pc in approximately 73,000
years.)  If we continue this exercise, looking out the rearview
mirror, so to speak, it should be possible to reconstruct a
deterministic history of the ISM that the Sun experienced in the
not-too-distant past.  The Sun's ISM history could then be converted
into a cosmic ray flux history, based on the heliospheric response to
the historical interstellar density profile.  As we sample more
distant environments, and therefore more distant times, the peculiar
motions of the Sun and the ISM cease to make a true deterministic
history of the Earth's cosmic ray flux possible, but would still
represent a possible and plausible cosmic ray history of the Earth, or
any other planet orbiting a nearby star.  Although the most recently
experienced ISM environments may not include dense molecular clouds,
nor the most recent climatic history include dramatic periods of
global glaciations, such a short-term test does not suffer the
systematics that make long-term correlations difficult.  Both the
astronomical ({\it Hipparcos} distances to nearby stars) and climatic
\citep{zachos01} records of the most recent past are sampled at high
temporal resolution and are well calibrated.

\section{Conclusions}

The LISM is a unique conduit that connects large scale galactic and
extragalactic structures with planetary atmospheres.  Although every
observation of an astronomical object outside our solar system (and
even some ``within'' our solar system) peers through the LISM, we do
not, as of yet, have a detailed three-dimensional global understanding
of the morphological or physical properties of our own interstellar
environment.  The challenges of observing the properties of the LISM
are slowly being met with larger databases of high resolution spectra
taken from the FUV to the optical, along sightlines toward nearby
stars.  Due to the limited number of transitions that are sensitive to
the low column densities of LISM material, it is critical to have
access to high resolution spectrographs from the FUV to the optical.
With the recent loss of the Space Telescope Imaging Spectrograph
(STIS) on {\it HST}, no high resolution astronomical spectrograph
($R\,\equiv\,\lambda/\Delta\lambda\,\geq\,$50,000) is currently
operating in the ultraviolet.  A new high resolution UV instrument is
needed in order to preserve and expand our observational capabilities
in the UV.  Without such a facility, among other losses, we will no
longer have the ability to observe gas in the LISM, and risk being
ignorant of our most immediate interstellar surroundings.  The LISM
interacts with stellar and planetary systems and as we explore the
subtle astrobiological parameters that control the degrees of
habitability of planets, the role of the LISM may turn out to be
significant.

Since the LISM, and the ISM in general, is an important part of a
grand ``galactic ecology,'' research on the LISM touches many
different areas of astrophysics, often at a profound level.  Carl
Sagan once said that ``we are made of star-stuff'' \citep{sagan80}.
Created in stars, our ``stuff'' was mixed and transported from across
the galaxy by the ISM and, eventually, it is likely that this
``stuff'' will return there to be recycled yet again.

\acknowledgements  
Support for this work was provided by NASA through Hubble Fellowship grant  HST-HF-01190.01 awarded by the Space Telescope Science Institute, which is operated by the Association of Universities for Research in Astronomy, Inc., for NASA, under contract NAS 5-26555.  I would like to thank John Scalo and Brian Wood for their helpful suggestions.

%\bibliographystyle{apj}
%\bibliography{refs}

\vspace{-2mm}
\begin{thebibliography}{}
\expandafter\ifx\csname natexlab\endcsname\relax\def\natexlab#1{#1}\fi

\bibitem[{{Alpen}(1998)}]{alpen98}
{Alpen}, E.~L. 1998, {Radiation Biophysics}, 2nd edn. (San Diego: Academic
  Press)

\bibitem[{{Asplund} {et~al.}(2005){Asplund}, {Grevesse}, \&
  {Sauval}}]{asplund05}
{Asplund}, M., {Grevesse}, N., \& {Sauval}, A.~J. 2005, in ASP Conf. Ser. 336:
  Cosmic Abundances as Records of Stellar Evolution and Nucleosynthesis in
  honor of David L. Lambert, ed. T.~G. {Barnes} \& F.~N. {Bash} (San Francisco:
  ASP), 25

\bibitem[{{Baranov}(1990)}]{baranov90}
{Baranov}, V.~B. 1990, Space Science Reviews, 52, 89

\bibitem[{{Begelman} \& {Rees}(1976)}]{begelman76}
{Begelman}, M.~C., \& {Rees}, M.~J. 1976, \nat, 261, 298

%\bibitem[{{Breitschwerdt} \& {Cox}(2004)}]{breitschwerdt04}
%{Breitschwerdt}, D., \& {Cox}, D.~P. 2004, in ASSL Vol. 315: How Does the
%  Galaxy Work?, ed. E.~J. {Alfaro} (Dordrecht: Kluwer Academic Publishers), 391

\bibitem[{{Brownlee} {et~al.}(2003)}]{brownlee03}
{Brownlee}, D.~E., {et~al.} 2003, J. Geophysical Research (Planets), 108, 1

\bibitem[{{Burton}(2004)}]{burton04}
{Burton}, M. 2004, in IAU Symp. 213, Bioastronomy 2002: Life Among the Stars,
  ed. R.~{Norris} \& F.~{Stootman} (San Francisco: ASP), 123

\bibitem[{{Carslaw} {et~al.}(2002){Carslaw}, {Harrison}, \&
  {Kirkby}}]{carslaw02}
{Carslaw}, K.~S., {Harrison}, R.~G., \& {Kirkby}, J. 2002, Science, 298, 1732

\bibitem[{{Chol Minh} {et~al.}(2003)}]{cholminh03}
{Chol Minh}, Y.~C.~Y., {et~al.} 2003, New Astronomy, 8, 795

\bibitem[{{Cowie} \& {Songaila}(1986)}]{cowie86}
{Cowie}, L.~L., \& {Songaila}, A. 1986, \araa, 24, 499

%\bibitem[{{Cox}(2005)}]{cox05}
%{Cox}, D.~P. 2005, \araa, 43, 337

\bibitem[{{Crawford}(2001)}]{crawford01}
{Crawford}, I.~A. 2001, \mnras, 327, 841

\bibitem[{{Dehnen} \& {Binney}(1998)}]{dehnen98}
{Dehnen}, W., \& {Binney}, J.~J. 1998, \mnras, 298, 387

\bibitem[{{Dickey} \& {Lockman}(1990)}]{dickey90}
{Dickey}, J.~M., \& {Lockman}, F.~J. 1990, \araa, 28, 215

\bibitem[{{Dupuis} {et~al.}(1995){Dupuis}, {Vennes}, {Bowyer}, {Pradhan}, \&
  {Thejll}}]{dupuis95}
{Dupuis}, J., {Vennes}, S., {Bowyer}, S., {Pradhan}, A.~K., \& {Thejll}, P.
  1995, \apj, 455, 574

\bibitem[{{Fahr}(1968)}]{fahr68}
{Fahr}, H.~J. 1968, \apss, 2, 474

\bibitem[{{Ferlet}(1999)}]{ferlet99}
{Ferlet}, R. 1999, \aapr, 9, 153

\bibitem[{{Florinski} {et~al.}(2004){Florinski}, {Pogorelov}, {Zank}, {Wood},
  \& {Cox}}]{florinski04}
{Florinski}, V., {Pogorelov}, N., {Zank}, G., {Wood}, B.~E., \& {Cox}, D.~P.
  2004, \apj, 604, 700

\bibitem[{{Florinski} {et~al.}(2003){Florinski}, {Zank}, \&
  {Axford}}]{florinski03}
{Florinski}, V., {Zank}, G.~P., \& {Axford}, W.~I. 2003, Geophys. Res. Lett.,
  30, 5

\bibitem[{{Frisch}(1995)}]{frisch95}
{Frisch}, P.~C. 1995, Space Science Reviews, 72, 499

\bibitem[{{Frisch}(1998)}]{frisch98}
{Frisch}, P.~C. 1998, in Planetary systems: the long view, ed. L.~M.
  {Celnikier} \& J.~{Tran Than Van}, 1

\bibitem[{{Frisch}(2004)}]{frisch04}
{Frisch}, P.~C. 2004, Advances in Space Research, 34, 20

\bibitem[{{Frisch} {et~al.}(2002){Frisch}, {Grodnicki}, \& {Welty}}]{frisch02}
{Frisch}, P.~C., {Grodnicki}, L., \& {Welty}, D.~E. 2002, \apj, 574, 834

\bibitem[{{Gies} \& {Helsel}(2005)}]{gies05}
{Gies}, D.~R., \& {Helsel}, J.~W. 2005, \apj, 626, 844

\bibitem[{{Gloeckler} {et~al.}(1997){Gloeckler}, {Fisk}, \&
  {Geiss}}]{gloeckler97}
{Gloeckler}, G., {Fisk}, L.~A., \& {Geiss}, J. 1997, \nat, 386, 374

\bibitem[{{Gloeckler} {et~al.}(2004)}]{gloeckler04}
{Gloeckler}, G., {et~al.} 2004, \aap, 426, 845

\bibitem[{{Gurevich} \& {Zybin}(2001)}]{gurevich01}
{Gurevich}, A.~V., \& {Zybin}, K.~P. 2001, Uspekhi Fizicheskikh Nauk, 44, 1119

\bibitem[{{Hoffman} {et~al.}(1998){Hoffman}, {Kaufman}, {Halverson}, \&
  {Schrag}}]{hoffman98}
{Hoffman}, P.~F., {Kaufman}, A.~J., {Halverson}, G., \& {Schrag}, D. 1998,
  Science, 281, 1342

\bibitem[{{Hoyle} \& {Lyttleton}(1939)}]{hoyle39}
{Hoyle}, F., \& {Lyttleton}, R.~A. 1939, in Proc. of the Cambridge
  Philisophical Society, 405

\bibitem[{{Hurwitz} {et~al.}(2005){Hurwitz}, {Sasseen}, \& {Sirk}}]{hurwitz05}
{Hurwitz}, M., {Sasseen}, T.~P., \& {Sirk}, M.~M. 2005, \apj, 623, 911

\bibitem[{{Jenkins}(2002)}]{jenkins02}
{Jenkins}, E.~B. 2002, \apj, 580, 938

\bibitem[{{Jenkins} {et~al.}(2000)}]{jenkins00}
{Jenkins}, E.~B., {et~al.} 2000, \apjl, 538, L81

\bibitem[{{Krimigis} {et~al.}(2003)}]{krimigis03}
{Krimigis}, S.~M., {et~al.} 2003, \nat, 426, 45

\bibitem[{{Kurth} \& {Gurnett}(2003)}]{kurth03}
{Kurth}, W.~S., \& {Gurnett}, D.~A. 2003, J. Geophysical Research (Space
  Physics), 108, 2

\bibitem[{{Lallement}(2004)}]{lallement04}
{Lallement}, R. 2004, \aap, 418, 143

\bibitem[{{Lallement} \& {Bertin}(1992)}]{lallement92}
{Lallement}, R., \& {Bertin}, P. 1992, \aap, 266, 479

\bibitem[{{Lallement} {et~al.}(2005){Lallement}, {Qu{\'e}merais}, {Bertaux},
  {Ferron}, {Koutroumpa}, \& {Pellinen}}]{lallement05}
{Lallement}, R., {Qu{\'e}merais}, E., {Bertaux}, J.~L., {Ferron}, S.,
  {Koutroumpa}, D., \& {Pellinen}, R. 2005, Science, 307, 1447

\bibitem[{{Lallement} {et~al.}(1986){Lallement}, {Vidal-Madjar}, \&
  {Ferlet}}]{lallement86}
{Lallement}, R., {Vidal-Madjar}, A., \& {Ferlet}, R. 1986, \aap, 168, 225

\bibitem[{{Lallement} {et~al.}(2003){Lallement}, {Welsh}, {Vergely}, {Crifo},
  \& {Sfeir}}]{lallement03}
{Lallement}, R., {Welsh}, B.~Y., {Vergely}, J.~L., {Crifo}, F., \& {Sfeir}, D.
  2003, \aap, 411, 447

\bibitem[{{Lehner} {et~al.}(2003){Lehner}, {Jenkins}, {Gry}, {Moos}, {Chayer},
  \& {Lacour}}]{lehner03}
{Lehner}, N., {Jenkins}, E.~B., {Gry}, C., {Moos}, H.~W., {Chayer}, P., \&
  {Lacour}, S. 2003, \apj, 595, 858

\bibitem[{{Linsky} \& {Wood}(1996)}]{linsky96}
{Linsky}, J.~L., \& {Wood}, B.~E. 1996, \apj, 463, 254

\bibitem[{{Lockman}(2002)}]{lockman02}
{Lockman}, F.~J. 2002, in ASP Conf. Ser. 276: Seeing Through the Dust, ed.
  A.~R. {Taylor}, T.~L. {Landecker}, \& A.~G. {Willis} (San Francisco: ASP),
  107

\bibitem[{{Magnani} {et~al.}(1996){Magnani}, {Hartmann}, \&
  {Speck}}]{magnani96}
{Magnani}, L., {Hartmann}, D., \& {Speck}, B.~G. 1996, \apjs, 106, 447

\bibitem[{{Marsh} \& {Svensmark}(2000)}]{marsh00}
{Marsh}, N.~D., \& {Svensmark}, H. 2000, Physical Review Letters, 85, 5004

\bibitem[{{McCray} \& {Kafatos}(1987)}]{mccray87}
{McCray}, R., \& {Kafatos}, M. 1987, \apj, 317, 190

\bibitem[{{McDonald} {et~al.}(2003){McDonald}, {Stone}, {Cummings}, {Heikkila},
  {Lal}, \& {Webber}}]{mcdonald03}
{McDonald}, F.~B., {Stone}, E.~C., {Cummings}, A.~C., {Heikkila}, B., {Lal},
  N., \& {Webber}, W.~R. 2003, \nat, 426, 48

\bibitem[{{McKay} \& {Thomas}(1978)}]{mckay78}
{McKay}, C.~P., \& {Thomas}, G.~E. 1978, Geophys. Res. Lett., 5, 215

\bibitem[{{Mewaldt} \& {Liewer}(2001)}]{mewaldt01}
{Mewaldt}, R.~A., \& {Liewer}, P.~C. 2001, in COSPAR Colloq. Ser. 11, The Outer
  Heliosphere: The Next Frontiers, ed. K.~{Scherer}, H.~{Fichtner}, H.~J.
  {Fahr}, \& E.~{Marsch} (Amsterdam: Pergamon Press), 451

\bibitem[{{M{\"o}bius} {et~al.}(2004)}]{mobius04}
{M{\"o}bius}, E., {et~al.} 2004, \aap, 426, 897

\bibitem[{{Neckel} {et~al.}(1980){Neckel}, {Klare}, \& {Sarcander}}]{neckel80}
{Neckel}, T., {Klare}, G., \& {Sarcander}, M. 1980, \aaps, 42, 251

\bibitem[{{Oegerle} {et~al.}(2005){Oegerle}, {Jenkins}, {Shelton}, {Bowen}, \&
  {Chayer}}]{oegerle05}
{Oegerle}, W., {Jenkins}, E., {Shelton}, R., {Bowen}, D., \& {Chayer}, P. 2005,
  \apj, 622, 377

\bibitem[{{Pavlov} {et~al.}(2005){Pavlov}, {Toon}, {Pavlov}, {Bally}, \&
  {Pollard}}]{pavlov05}
{Pavlov}, A.~A., {Toon}, O.~B., {Pavlov}, A.~K., {Bally}, J., \& {Pollard}, D.
  2005, Geophys. Res. Lett., 32, 3705

\bibitem[{{Pogorelov} {et~al.}(2004){Pogorelov}, {Zank}, \&
  {Ogino}}]{pogorelov04}
{Pogorelov}, N.~V., {Zank}, G.~P., \& {Ogino}, T. 2004, \apj, 614, 1007

%\bibitem[{{Rand} \& {Kulkarni}(1989)}]{rand89}
%{Rand}, R.~J., \& {Kulkarni}, S.~R. 1989, \apj, 343, 760

\bibitem[{{Rand} \& {Lyne}(1994)}]{rand94}
{Rand}, R.~J., \& {Lyne}, A.~G. 1994, \mnras, 268, 497

\bibitem[{{Randall} {et~al.}(2005)}]{randall05}
{Randall}, C.~E., {et~al.} 2005, Geophys. Res. Lett., 32, 5802

\bibitem[{{Rauch} {et~al.}(1999){Rauch}, {Sargent}, \& {Barlow}}]{rauch99}
{Rauch}, M., {Sargent}, W.~L.~W., \& {Barlow}, T.~A. 1999, \apj, 515, 500

\bibitem[{{Redfield} \& {Linsky}(2000)}]{redfield00}
{Redfield}, S., \& {Linsky}, J.~L. 2000, \apj, 534, 825

\bibitem[{{Redfield} \& {Linsky}(2001)}]{redfield01}
{Redfield}, S., \& {Linsky}, J.~L. 2001, \apj, 551, 413

\bibitem[{{Redfield} \& {Linsky}(2002)}]{redfield02}
{Redfield}, S., \& {Linsky}, J.~L. 2002, \apjs, 139, 439

\bibitem[{{Redfield} \& {Linsky}(2004{\natexlab{a}})}]{redfield04sw}
{Redfield}, S., \& {Linsky}, J.~L. 2004{\natexlab{a}}, \apj, 602, 776

\bibitem[{{Redfield} \& {Linsky}(2004{\natexlab{b}})}]{redfield04tt}
{Redfield}, S., \& {Linsky}, J.~L. 2004{\natexlab{b}}, \apj, 613, 1004

\bibitem[{{Reedy} {et~al.}(1983){Reedy}, {Arnold}, \& {Lal}}]{reedy83}
{Reedy}, R.~C., {Arnold}, J.~R., \& {Lal}, D. 1983, Science, 219, 127

\bibitem[{{Richardson}(1997)}]{richardson97}
{Richardson}, J.~D. 1997, Geophys. Res. Lett., 24, 2889

\bibitem[{{Sagan}(1980)}]{sagan80}
{Sagan}, C. 1980, {Cosmos} (New York: Random House)

\bibitem[{{Savage} \& {Sembach}(1996)}]{savage96}
{Savage}, B.~D., \& {Sembach}, K.~R. 1996, \araa, 34, 279

\bibitem[{{Shapley}(1921)}]{shapley21}
{Shapley}, H. 1921, J. Geology, 29, 502

\bibitem[{{Shaviv}(2003)}]{shaviv03}
{Shaviv}, N.~J. 2003, New Astronomy, 8, 39

\bibitem[{{Slavin} \& {Frisch}(2002)}]{slavin02}
{Slavin}, J.~D., \& {Frisch}, P.~C. 2002, \apj, 565, 364

\bibitem[{{Snowden} {et~al.}(1998){Snowden}, {Egger}, {Finkbeiner}, {Freyberg},
  \& {Plucinsky}}]{snowden98}
{Snowden}, S.~L., {Egger}, R., {Finkbeiner}, D.~P., {Freyberg}, M.~J., \&
  {Plucinsky}, P.~P. 1998, \apj, 493, 715

\bibitem[{{Stone} {et~al.}(2005){Stone}, {Cummings}, {McDonald}, {Heikkila},
  {Lal}, \& {Webber}}]{stone05}
{Stone}, E.~C., {Cummings}, A.~C., {McDonald}, F.~B., {Heikkila}, B.~C., {Lal},
  N., \& {Webber}, W.~R. 2005, Science, 309, 2017

\bibitem[{{Svensmark} \& {Friis-Christensen}(1997)}]{svensmark97}
{Svensmark}, H., \& {Friis-Christensen}, E. 1997, J. Atmos. and Terrestrial
  Phys., 59, 1225

\bibitem[{{Thaddeus}(1986)}]{thaddeus86}
{Thaddeus}, P. 1986, The Galaxy and the Solar System, 61

\bibitem[{{Tinbergen}(1982)}]{tinbergen82}
{Tinbergen}, J. 1982, \aap, 105, 53

\bibitem[{{Vallerga} {et~al.}(1993){Vallerga}, {Vedder}, {Craig}, \&
  {Welsh}}]{vallerga93}
{Vallerga}, J.~V., {Vedder}, P.~W., {Craig}, N., \& {Welsh}, B.~Y. 1993, \apj,
  411, 729

\bibitem[{{Wallmann}(2004)}]{wallmann04}
{Wallmann}, K. 2004, Geochemistry, Geophysics, Geosystems, 5, 6004

\bibitem[{{Welty} \& {Hobbs}(2001)}]{welty01}
{Welty}, D.~E., \& {Hobbs}, L.~M. 2001, \apjs, 133, 345

\bibitem[{{Welty} {et~al.}(1994){Welty}, {Hobbs}, \& {Kulkarni}}]{welty94}
{Welty}, D.~E., {Hobbs}, L.~M., \& {Kulkarni}, V.~P. 1994, \apj, 436, 152

\bibitem[{{Welty} {et~al.}(1996){Welty}, {Morton}, \& {Hobbs}}]{welty96}
{Welty}, D.~E., {Morton}, D.~C., \& {Hobbs}, L.~M. 1996, \apjs, 106, 533

\bibitem[{{Witte}(2004)}]{witte04}
{Witte}, M. 2004, \aap, 426, 835

\bibitem[{{Wood}(2004)}]{wood04}
{Wood}, B.~E. 2004, Living Reviews in Solar Physics, 1, 2

\bibitem[{{Wood} \& {Linsky}(1997)}]{wood97}
{Wood}, B.~E., \& {Linsky}, J.~L. 1997, \apjl, 474, L39

\bibitem[{{Wood} {et~al.}(2001){Wood}, {Linsky}, {M{\"u}ller}, \&
  {Zank}}]{wood01}
{Wood}, B.~E., {Linsky}, J.~L., {M{\"u}ller}, H.-R., \& {Zank}, G.~P. 2001,
  \apjl, 547, L49

\bibitem[{{Wood} {et~al.}(2005{\natexlab{a}}){Wood}, {M{\"u}ller}, {Zank},
  {Linsky}, \& {Redfield}}]{wood05let}
{Wood}, B.~E., {M{\"u}ller}, H.-R., {Zank}, G., {Linsky}, J., \& {Redfield}, S.
  2005{\natexlab{a}}, \apjl, 628, L143

\bibitem[{{Wood} {et~al.}(2005{\natexlab{b}}){Wood}, {Redfield}, {Linsky},
  {M{\"u}ller}, \& {Zank}}]{wood05sup}
{Wood}, B.~E., {Redfield}, S., {Linsky}, J.~L., {M{\"u}ller}, H.-R., \& {Zank},
  G.~P. 2005{\natexlab{b}}, \apjs, 159, 118

\bibitem[{{Wood} {et~al.}(2002){Wood}, {Redfield}, {Linsky}, \&
  {Sahu}}]{wood02}
{Wood}, B.~E., {Redfield}, S., {Linsky}, J.~L., \& {Sahu}, M.~S. 2002, \apj,
  581, 1168

\bibitem[{{Yeghikyan} \& {Fahr}(2004)}]{yeghikyan04a}
{Yeghikyan}, A., \& {Fahr}, H. 2004, \aap, 415, 763

\bibitem[{{Zachos} {et~al.}(2001){Zachos}, {Pagani}, {Sloan}, {Thomas}, \&
  {Billups}}]{zachos01}
{Zachos}, J., {Pagani}, M., {Sloan}, L., {Thomas}, E., \& {Billups}, K. 2001,
  Science, 292, 686

\bibitem[{{Zank}(1999)}]{zank99}
{Zank}, G.~P. 1999, Space Science Reviews, 89, 413

\bibitem[{{Zank} \& {Frisch}(1999)}]{zankfrisch99}
{Zank}, G.~P., \& {Frisch}, P.~C. 1999, \apj, 518, 965

\end{thebibliography}

\end{document}